# Reconfiguration of a Magnetic Tunnel Junction as a Way to Turn It into a Field-Free Vortex Oscillator

Maksim Stebliy, Alex Jenkins, Luana Benetti, Ricardo Ferreira

INL - International Iberian Nanotechnology Laboratory, Braga 4715-330, Portugal
maksim.steblii@inl.int

## Abstract

*Magnetic tunnel junctions (MTJs) are key elements in practical spintronics, enabling not only conventional tasks such as data storage, transmission, and processing but also the implementation of compute-in-memory processing elements, facilitating the development of efficient hardware for neuromorphic computing. The functionality of an MTJ is determined by the properties of its free layer (FL) and reference layer (RL) with fixed magnetization, separated by an MgO tunnel barrier. This paper presents a mechanism for reconfiguring the RL, which is the upper layer of a pinned synthetic antiferromagnet, enabling a reversible transition from a single-domain state to a vortex magnetic state with different core positions. When the RL is in the vortex state, it generates a spin current with a vortex-like polarization distribution, enabling stable vortex oscillations in the FL even in the absence of external magnetic fields. This effect has been confirmed in MTJs with diameters ranging from 400 to 1000 nm. It is demonstrated, using experimental data with comparative micromagnetic simulation, that the pinning antiferromagnet retains a long term memory of previous reannealing states resulting in a deformation of the vortex polarised spin current, which in turn introduces a strong dynamical vortex core polarity symmetry breaking. The analogue reprogrammable nature of both the static and dynamic properties of the MTJ demonstrate different possible routes for the introduction of non-volatility into radiofrequency spintronic neuromorphic paradigms.*

## 1. Introduction

General-purpose technologies, such as artificial intelligence, are driving the need for compute-in-memory data processing elements. Spintronics is a promising direction in this field, with devices characterized by non-linearity, complex dynamics, hysteresis, non-volatility, and collective behaviour [1] [2]. Recent studies show that hardware implementations of neuromorphic cmputing systems using spintronic elements can offer compact and efficient solutions [3] [4]. These systems can perform simple tasks, such as threshold detection [5] [6], as well as more complex functions, including wave type, audio and drone classification [7] [8], or image and speech recognition [9] [10] [11].

The fundamental element of spintronics is the magnetic tunnel junction (MTJ), which consists of a free layer (FL) and a reference layer (RL) separated by an MgO barrier [12]. This structure converts changes in the FL magnetic state into resistance variations via tunnel magnetoresistance (TMR), which can reach several hundred percent [13]. Based on this principle, various devices can be implemented, including memory cells, field sensors, logic elements [14], spin-torque diodes [15] [16], and spin-torque nano-oscillators [17]. Devices of the last category have applications in energy harvesting [18] [19], wireless communication [20] [21], spectrum analysis [22], random number generation [23] [24], and artificial neuron [25-28]. Oscillations arise from the interaction of current with various magnetic structures, such as single-domain states [25] [26], skyrmions [27] [28], droplets [29], and vortices [30] [31] [32] [33, 33]. Among these, vortices exhibit the lowest frequency, on the order of a few hundred megahertz, but offer the highest power and narrowest linewidth of oscillation [17].

The vortex structure is characterized by in-plane magnetization rotation and out-of-plane magnetization in the central core. It is defined by two topological numbers: chirality (±C), indicating the rotation direction (clockwise or counterclockwise), and polarity (±P), representing the core magnetization orientation (up or down). The core acts as a quasiparticle, shifting its position under an external magnetic field [34] or spin-polarized current [35].

Previously, vortex spin-torque nano-oscillators were used to implement radio frequency neurons, leveraging a device's ability to convert a DC signal into an AC signal once the input threshold is exceeded [36] [37] [38]. In order to generate steady state oscillations a perpendicularly polarised current is required, which has previously been achieved using a perpendicular polariser [31] [39], but is most commonly achieved by applying a large (~0.5 T) perpendicular magnetic field in order to tilt the polariser out-of-plane [30] [8]. In this work, using a standard MTJ structure, we experimentally demonstrate the possibility of achieving a radio-frequency neuron using a reprogrammable vortex polarised spin current capable of achieving zero field steady state oscillations. Auto-oscillations induced by a vortex polarised spin current have been previously investigated theoretically [40] [41] [42] and observed experimentally [43] [44] [45] [46] in spin valves where an unpinned vortex was used as polariser, but systematic study was difficult due to the dynamic behaviour of the reference layer.

In the MTJ, the RL acts as a polarizer, setting the spin current polarization and influencing the FL magnetization through spin-transfer torque (STT) [47] [48]. The reprogrammable polariser in this work is made possible by locally annealing the MTJ device via Joule heating to a temperature above the Néel temperature of the pinning antiferromagnet, thus generating a magnetic vortex in the reference layer whose magnetisation is fixed. By modifying the conditions during local annealing process, the position of the fixed vortex core can be precisely controlled, however even after reannealing the antiferromagnet retains a long term memory in the form of a deformation along a historical axis. The deformation of the antiferromagnet introduces a clear anisotropy to the resistance dependence, and also has a strong impact on the dynamics of the free layer, with a clear symmetry dependence emerging with relation to the vortex core polarity (i.e. the sign of the out of plane vortex core).

In this work we demonstrate how the resistance and steady state oscillations of a magnetic vortex radiofrequency neuron can be reprogrammed both in an analogue manner due to a Joule heating induced realignment of the antiferromagnet as well as the binary internal degrees of freedom in a magnetic vortex (i.e. chirality and polarity).

## 2. Description of the Device and Concepts of Use

Experimental studies were conducted on magnetic tunnel junctions with pillar diameters ranging from 100 to 1000 nm with the composition IrMn(6)/CoFe(2)/Ru(0.7)/CoFeB(2.6)/MgO/CoFeB(2)/Ta(0.2)/NiFe(7 nm). The full stack and preparation details are provided in the *Materials and Methods* section, while and the magnetic parameters are given in S.1 section. The MgO layer separates the free layer (FL) from the synthetic antiferromagnetic structure (SAF) and contributes 140% of the TMR, with a resistance-to-area ratio of 5 Ω·µm². The SAF consists of reference (RL) and pinning (PL) layers, which exhibit a strictly antiparallel magnetization alignment due to indirect exchange coupling and remain fixed as a result of exchange bias, Fig. 1a.

It was found that applying a voltage pulse to the MTJ enables reconfiguration of the fixed magnetic structure of the reference layer. The reconfiguration mechanism involves unlocking the SAF magnetization through Joule heating of the IrMn layer above the Néel temperature, followed by relaxation in the presence of external and Oersted fields. At the end of the pulse, the current magnetic state becomes fixed due to exchange bias as the IrMn layer cools. The described mechanism was used in our previous work to reconfigure the additional storage layer, enabling variation of the vortex resonance frequency in the free layer by adjusting the magnetostatic field [49].

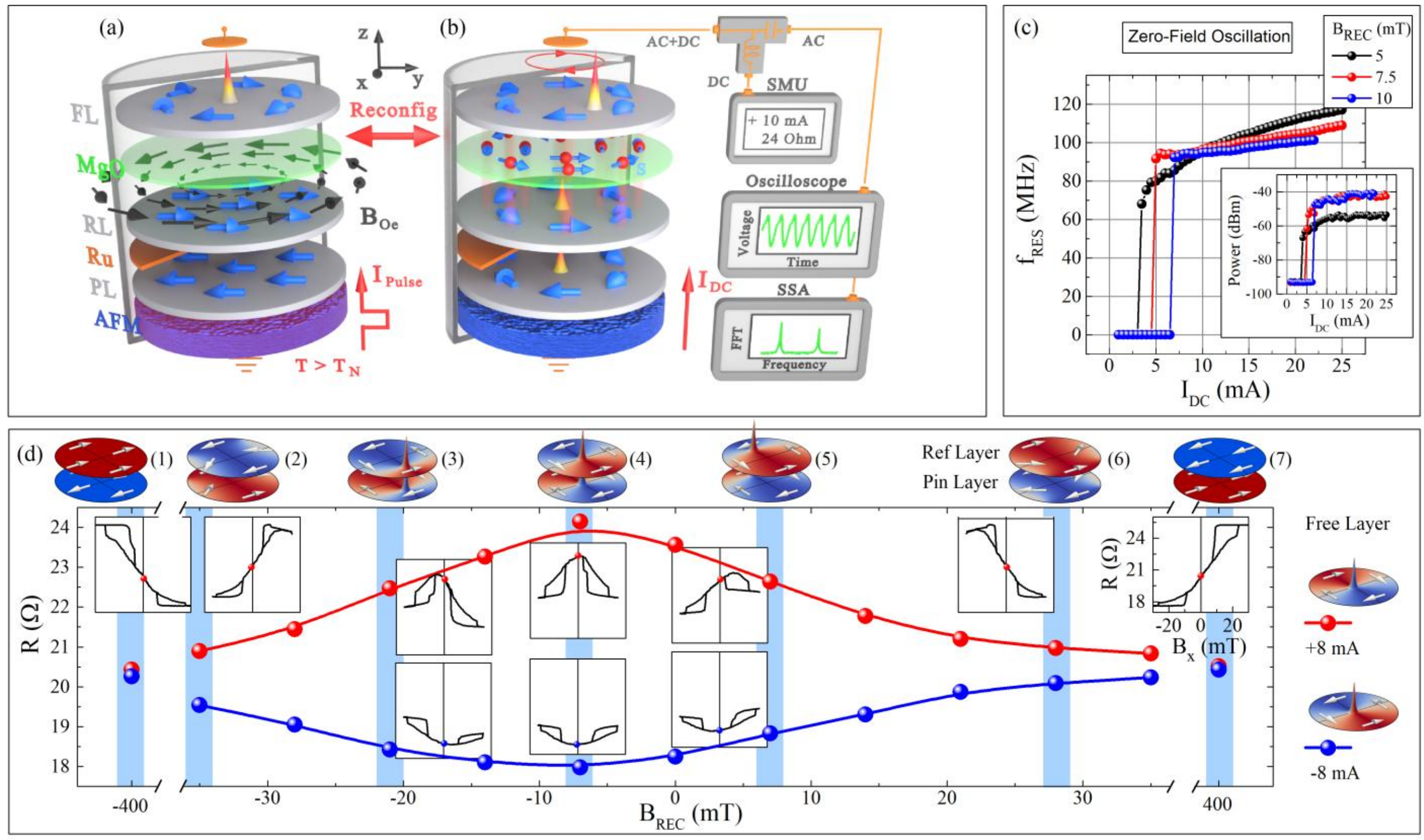

*Fig. 1. Schematic diagram of the MTJ structure in the single-domain (a) and vortex (b) states within the reference and pinning layers. (c) Dependence of the gyrotropic frequency of zero-field auto-oscillations on the current for different values of the reconfiguration field (diameter: 1000 nm). The insets show the corresponding dependencies of the reflected oscillation power. (d) Dependence of MTJ resistance on the magnetic state in the SAF in the absence of an external field, obtained under different reconfiguration fields. The red and blue curves correspond to cases where the vortex chirality in the free and reference layers is the different and same, respectively. The insets show the magnetic hysteresis loops for the corresponding SAF states.*

The pulse amplitude required for magnetization unlocking increases with pillar diameter, from 0.85 V for 100 nm to 1 V for 1000 nm. The Néel temperature was estimated to be 220°C for a continuous film with the same composition (Section S.2), though it may decrease with patterning [49]. The reconfiguration mechanism can operate with approximately 20 ns pulse duration, but 1 ms pulses were used in the presented results to simplify the circuit diagram.

## 3. Results

Section 1 examines the magnetic configurations of the SAF that result from reconfiguration and identifies cases where a vortex forms, supporting auto-oscillations. Section 2 analyzes how the polarity of the vortex in the FL layer affects the ability to achieve oscillations. Section 3 explores the relationship between the polarity dependence of oscillations and the conditions of reconfiguration. Finally, Section 4 provides a qualitative explanation of the auto-oscillation mechanism, which arises from the interaction between a magnetic vortex in the FL layer and a spin current with non-uniform polarization. All results presented in the main text of the article were obtained for MTJ with a diameter of 1000 nm. Results for smaller diameters are given in section 1.

### 3.1. Reconfiguration

Applying ±400 mT during reconfiguration is sufficient to overcome indirect exchange and align the magnetization in the PL and RL parallel. At the end of the pulse, the PL remains fixed, while the RL reverses to an antiparallel state when the field is switched off. As a result, the SAF is set into a single-domain state, as illustrated in States 1 and 7 in Fig. 1c, where the corresponding hysteresis loops, obtained by recording MTJ resistance, exhibit features characteristic of the nucleation, displacement, and annihilation of a magnetic vortex in the FL [34]. In these cases, the resistance at zero field has a value roughly half way between the value for the P and AP states, and is independent of the vortex chirality in the FL. The chirality of the vortex in the free layer is determined during the vortex nucleation process by the direction of DC current applied to the MTJ and the subsequent rotation of the Oersted field, as shown in Fig. 1a. For a device with a 1000 nm diameter, the minimal DC current sufficient for reliable chirality control is 8 mA, inducing 4 mT at the periphery (S.3).

Reconfiguration in a ±30 mT field is not sufficient to overcome the RKKY interaction in the SAF. As a result, the magnetization in the RL, due to its higher magnetic moment, aligns with the external field, while in the PL, it aligns oppositely. This ordering remains stable after the pulse and field are switched off. Since the Oersted field is not negligible compared to the external field, the magnetic structure is no longer a single domain and shows signs of curvature, States 2 and 6 in Fig. 1c. Magnetization curvature in the RL makes the residual resistance dependent on the vortex chirality in the FL.

The voltage pulse in the absence of an external field results in the fixation of a pair of magnetic vortexes with opposite chirality in the SAF, as shown in State 4 of Fig. 1c. The magnetic structure of the RL follows the chirality of the Oersted field that can reach 23 mT at the periphery during the pulse (S.3). The PL aligns oppositely to remain strictly antiparallel due to indirect exchange with an energy of -0.3 mJ/m² (S.1). The corresponding hysteresis loops differ qualitatively depending on the vortex chirality in the free layer. High and low resistance states are observed at zero field for antiparallel and parallel combination of vortex in the FL and RL, respectively.

Reconfiguration in the ±20 mT range of field establishes a vortex structure in the SAF with a shifted vortex core, as depicted in States 3 and 5 of Fig. 1c. Corresponding hysteresis loops have non equal resistance at saturation in positive and negative sweeping field.

The reconfiguration of the SAF enables a gradual change in the residual resistance of the MTJ, splitting the dependence based on the chirality of the vortex in the FL. Additionally, it was found that near states 3 and 5, the DC current can excite steady-state gyrotropic oscillations of the vortex core in the FL. Auto-oscillations were detected using a serial spectrum analyzer (SSA) and an oscilloscope, both connected through a bias-tee, Fig.1b.

Steady-state oscillations are observed in the high-resistance state when the vortices in the FL and RL have opposite chirality and the current direction is +I. Oscillations at -I current occur unsystematically and extremely rarely. Reconfiguration near state (5), obtained at a field of +7.5 mT, allows for a smooth adjustment of the critical current required to induce oscillations (Fig. 1c). This relationship can be used to tune the activation level when implementing an artificial neuron. Increasing the transmitted current leads to a nonlinear increase in frequency from 70 to 120 MHz. In this case, the reflected signal power changes from -70 to -40 dBm (insert in Fig. 1c). The effect of zero-field auto-oscillation was also found in the MTJ with diameters of 800, 600, and 400 nm, S.5. Despite the fact that reconfiguration is able to work for diameters up to 100 nm, it was not possible to create a vortex in the SAF and, consequently, to induce oscillations in pillars smaller than 400 nm.

### 3.2. Conditions for the Excitation of Auto-Oscillations

To determine the relationship between the SAF state and the ability to induce auto-oscillation, two-dimensional (2D) field dependences were investigated. Within the studied field range, the magnetic structure of the RL remains fixed, while the vortex core in the FL can move continuously within the plane. The external field values along the x-axis ($B_x$) and y-axis ($B_y$) were increased in a spiral profile from zero. For each pair of field values at a fixed current of +8 mA, the resistance was measured simultaneously with the maximum power from the spectrum analyser to detect the presence of oscillations.

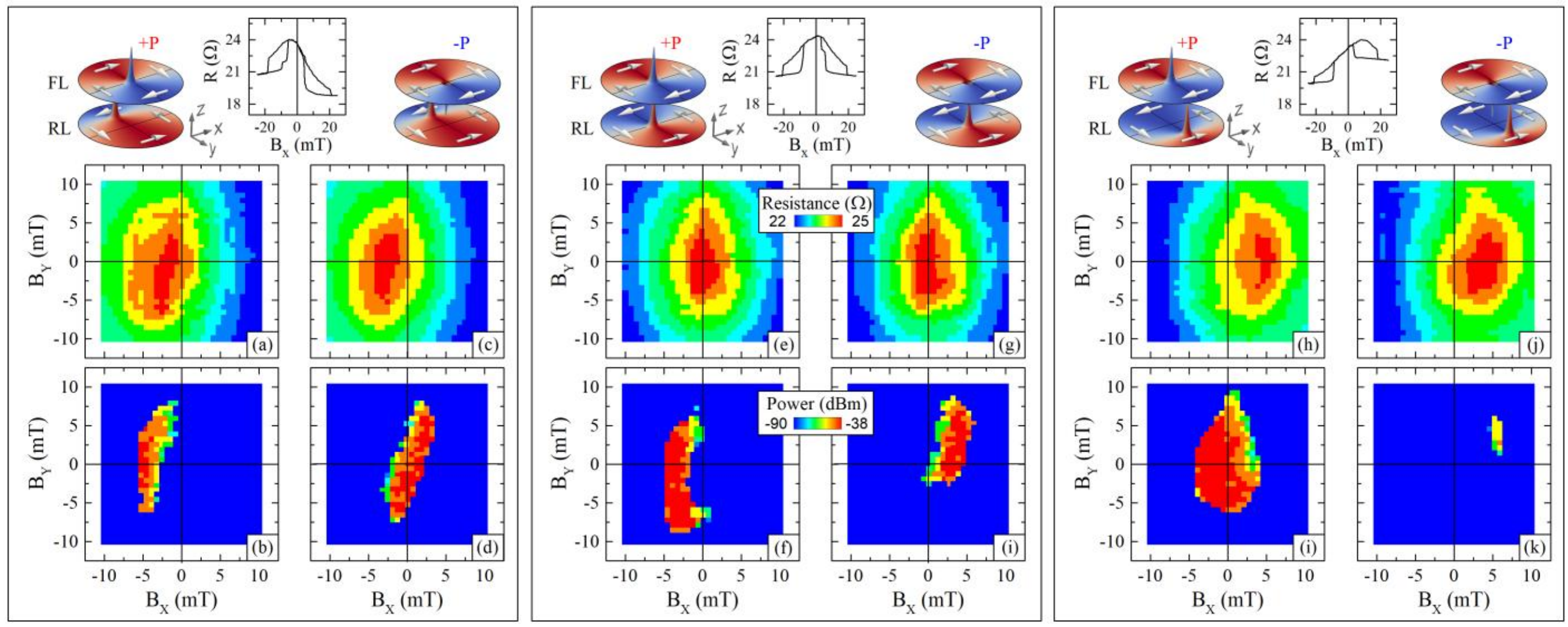

*Fig.2. Distribution of MTJ resistance and oscillation power as a function of the $B_x$ and $B_y$ fields, obtained for different polarities of the vortex core in the free layer (i.e., +P, -P). In all cases, a constant current of +8 mA was applied. Diagrams (a-d) correspond to the vortex core in the reference layer being shifted in the -y direction, diagrams (e-i) to the core being fixed at the center, and diagrams (h-k) to the core being shifted in the +y direction. The inset shows the hysteresis loop for the corresponding cases of the field sweep along the x-axis.*

In the first step, the values of the nonzero external field applied during reconfiguration were chosen to fix the vortex core in the SAF approximately at the centre. The corresponding resistance distribution map shows a maximum at the centre, where vortices with opposite chirality are positioned on top of each other, Fig.2e. The region of the in-plane $B_x$ and $B_y$ fields where oscillations can be observed is shifted to the left, Fig.2f.

Repeating this measurement after vortex re-nucleation in the FL allowed to find a region of oscillation mirrored to the right (Fig. 2i). A comparison of the resistance distribution for these two cases shows no significant differences (Fig. 2e, g). This indicates that the vortex chirality in the FL and the fixed position of the vortex in the RL remain the same. The only parameter that can change without affecting resistance is the polarity of the vortex core in the FL. The vortex core polarity is determined during the nucleation process and is probabilistic. The presence of a symmetry axis for oscillations is discussed in Section 3, while the mechanism of polarity-dependent restriction is discussed in Section 4.

To achieve zero-field oscillations, the vortex in the SAF was shifted in the -y direction by applying $+B_x$ during reconfiguration. As a result, the resistance maximum shifted to the $-B_x$ region, as confirmed by the resistance map for both core polarities in the FL (Fig. 2a,c). Consequently, the region of oscillation shifted to the $-B_x$ in both cases (Fig. 2b, d). The vortex core with polarity conditionally down can oscillate in this case in the absence of external field.

The same effect can be achieved for the opposite polarity, provided that the vortex in the SAF is fixed with a shift toward -y (Fig. 2h,j). In this case, both oscillation regions shift to $+B_x$ (Fig. 2i,k). It is also possible to shift the oscillation region along $B_y$ field. The results of such reconfigurations are given in Section S.4.

### 3.3. Symmetry of Oscillation

As can be seen in Fig. 2e,g the resistance dependence on the in-plane field is not perfectly symmetric, where instead of the value in resistance decreasing decreasingly equally for $B_x$ and $B_y$, there is instead a shaper decrease for $B_x$ than By. This preferred axis is indicative of an anisotropy in the system, which is believed not to originate from the vortex in the free layer, due to the absence of stray fields and the weak interaction through the MgO layer (S.1), but rather in the vortex in the reference layer. This deformation in the reference layer will be shown to be responsible for the pronounced polarity dependence of the steady state oscillations.

The presence of a minimum current, due to the induction of the Oersted field, ensures that the lowest-energy state corresponds to the formation of vortices with opposite chirality in the pinning and reference layers (Section S.6). This occurs due to the antiferromagnetic indirect exchange and the higher saturation magnetization of layer the RL compared to the PL. As a result, the vortices are symmetrical, with their cores aligned directly beneath each other. It is therefore believed that the deformation of the reference layer comes indirectly via the coupling with the antiferromagnetic IrMn layer. During the reconfiguration process, the antiferromagnetic layer is cooled in the presence of exchange interactions with the magnetic structure of the PL layer, which becomes imprinted at the interface. However, the magnetic structure of the antiferromagnetic layer retains residual traces of the previous single-domain state obtained under strong fields. Since this could not be directly visualised, the following experiment was conducted to indirectly illustrate the assumption.

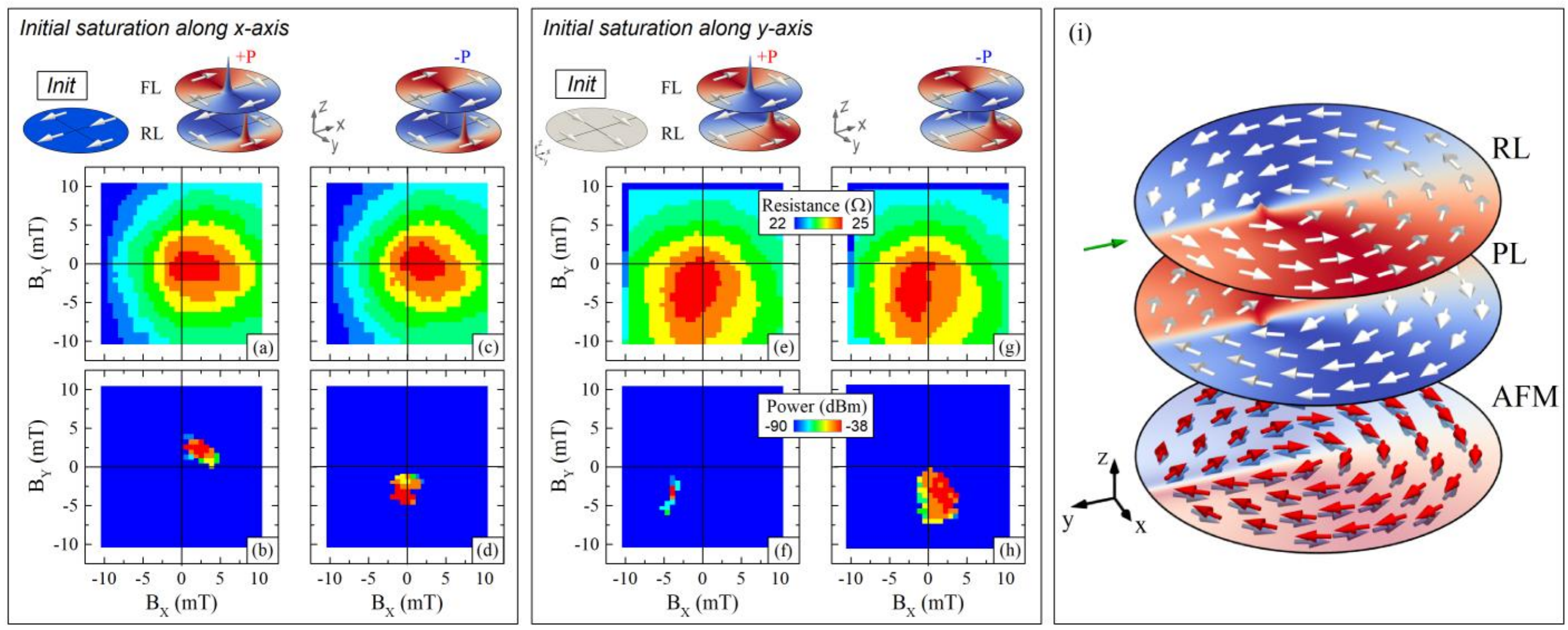

*Fig. 3. Distribution of the MTJ resistance (a, c) and oscillation power (b, d) as a function of the $B_x$ and $B_y$ fields, obtained for different vortex core polarities in the free layer. The magnetic structure fixed in the reference layer results from initial saturation along the x-axis, followed by reconfiguration in the absence of magnetic fields. Distributions (e, g) and (f, h) represent the corresponding resistance and oscillation power distributions for the case of initial initialization along the y-axis. (i) Schematic representation of the magnetic state of the IrMn antiferromagnetic layer, showing the deformation indicated by the green arrow and the resulting vortex displacement in the pinning and reference layers.*

The MTJ was reconfigured in the presence of a -400 mT field along the x-axis to establish a strictly single-domain state, similar to State 1 in Fig. 1c. The magnetic field was reduced to zero and a subsequent voltage pulse was applied and the two dimensional resistance map as a function to $B_x$ and $B_y$ was measured. The maximum of the resistance distribution is shifted along the x-axis and exhibits a deformation oriented along the x-axis (shown in Fig. 3a,c for both free layer vortex polarities). Similar to Fig. 3, the MTJ exhibits steady state oscillations (i.e. non zero rf power) for different magnetic field values depending upon the polarity of the vortex core in the free layer (Fig. 3b,d), with the different polarities being excited either side of symmetry axis orientated along the x direction.

To clarify the role of the initial state, the approach was repeated with the initial reconfiguration performed under a -400 mT field along the y-axis. Subsequent reconfiguration in zero field resulted in an orthogonal reversal of the symmetry axis of deformation on the resistance map (Fig. 3e,g). In this case, regions of oscillation were positioned on opposite sides of the y-axis (Fig. 3f,h).

Micromagnetic simulation was used to select the magnetic state of the antiferromagnetic layer, which, due to the effective exchange bias field, can induce uniaxial vortex deformations in the Pl and RL layers (Fig. 3i). A distinctive feature of the proposed antiferromagnetic structure is the absence of magnetic closure and the formation of a discontinuity line oriented along the initial saturation direction (indicated by an arrow). As shown below, the micromagnetic simulation results agree with the experiment in terms of both resistance distribution and polarity-dependent violation of oscillation symmetry.

### 3.4. The Mechanism of Vortex Auto-Oscillations

An electric current passing through the MTJ becomes spin-polarized as it passes through the RL layer and is subsequently injected into the FL layer. When the RL is fixed in the vortex state, the spin current's polarization repeats the vortex distribution. The interaction between the DC current with such non-uniform polarization and the magnetic vortex in the FL can give rise to steady-state oscillations [40]. A qualitative explanation of the mechanism of this effect can be obtained by considering the Slonczewski spin-transfer torque, which contains two components: the field-like torque ($\tau_{FL} \propto [m \times m_p]$) and the damping-like torque ($\tau_{DL} \propto \left[m \times [m_p \times m]\right]$), where $m_p$ is the spin current polarization and $m$ is the FL magnetization [46] [47]. The torques can be described in terms of effective fields: ($B_{FL} \propto m_p$) and ($B_{DL} \propto [m_p \times m]$), respectively. The first component is static, while the second can dynamically change with change of magnetisation in the FL. Using MuMax3 [47], it is possible to detect the effect of auto-oscillations using the standard function (S.8) and visualize the distribution of $B_{DL}$ using a custom field (S.9).

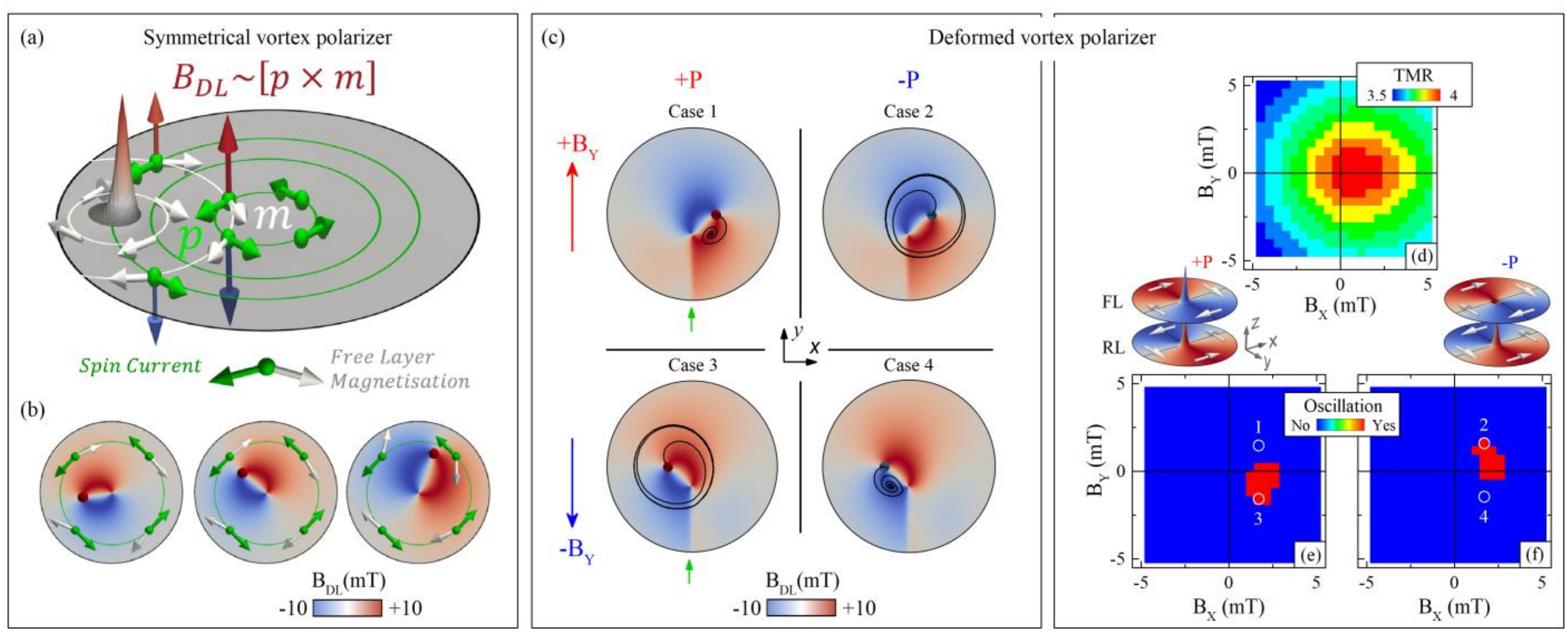

*Fig. 4. (a) Schematic representation of the effect of a spin current with a vortex polarization distribution on a magnetic vortex, resulting in the emergence of an effective $B_{DL}$ field due to spin-transfer torque. (b) Storyboard illustrating changes in the $B_{STT}$ field distribution as the vortex core position shifts, leading to stable gyrotropic motion. The current polarization*

*follows a symmetric vortex distribution. (c) The case of drop-shaped deformation of the current polarization vortex distribution. Diagrams (1–4) show the effective field distribution for different combinations of vortex core polarity and external field orientation at the initial moment after current activation. Black lines indicate the vortex core trajectory from its stationary state and the polarizer deformation indicated by the green arrow. (d) Corresponding simulation of the resistance distribution and (e, f) the presence of stable oscillations in an MTJ as a function of $B_x$ and $B_y$ fields for different vortex core polarities.*

First, the interaction of a magnetic vortex with a spin current having an undeformed vortex-like polarization is considered (Fig. 4a). The $B_{DL}$ field is defined by the vector cross product, which becomes nonzero for a non-collinear orientation between $m_p$ and m. To satisfy this condition, the vortex core in the FL must be displaced relative to the core in the RL. Since both vectors predominantly lie in the plane, $B_{DL}$ has only a perpendicular component, as illustrated in Fig. 4a. Consequently, the spin current with vortex polarization creates a local perpendicular field that attracts the magnetic vortex core in the FL.

The time evolution of this process is presented in Fig. 4b. As the vortex core moves, the effective field distribution will in turn change, resulting in continuous motion. Reversal of the core polarity changes the direction of rotation, but auto-oscillations occur in both cases under the same external field. The understanding of the perpendicular effective field which is produced by the spin current of a vortex polariser therefore can be explain the efficient dc to rf steady state oscillations observed in MTJs, but they cannot explain the pronounced polarity symmetry dependence which is observed.

Next, the interaction of a magnetic vortex with a spin current having a deformed vortex-like polarization, obtained by using the RL from Fig. 3i as a polarizer, is considered. The analysis starts with the case of the vortex core with polarity +P, displaced from the center by an external field $+B_x$ (Case 1 in Fig. 4c). When the current is applied, an effective field $B_{DL}$ is induced, with its initial distribution shown on the color map. A sharp field gradient appears in the region of polarizer deformation, marked by the green arrow. The current-induced motion of the vortex core from its equilibrium state is shown by the black line. The interaction with the deformation region attenuates the gyrotropic motion, making this region act as a potential barrier.

Changing the core polarity to the opposite −P under the same bias field $+B_x$ does not alter the effective field distribution (Case 2 in Fig. 4c). In this case, stable oscillations are observed. This behavior arises because polarity reversal changes the direction of the core's gyrotropic motion, allowing the vortex core to accelerate and gain sufficient kinetic energy to overcome the potential barrier. Reversing the external field direction to $-B_x$ results in auto-oscillations for polarity +P (Case 3), while oscillations are suppressed for polarity −P (Case 4).

The same simulations were carried out for various values of fields $B_x$ and $B_y$ to determine the presence of oscillations. Based on the results, oscillation diagrams were constructed for polarity +P (Fig. 4e) and polarity −P (Fig. 4f). Additionally, resistance maps were calculated (Fig. 4d) from the corresponding relaxed magnetic distributions for each field pair using the approach described in S.7. The simulation results qualitatively agree with the experimental data.

## 4. Discussion

The interaction between the magnetic vortex and the spin current of the symmetrical vortex polarization leads to oscillation of the vortex core in the free layer, regardless of its polarity and the external field. However, the experiment revealed oscillations only within a narrow range of fields for a fixed polarity. Based on the analysis of the experimental data, it is believed that the fixed vortex in the reference layer has a deformation along an axis determined by the initial saturation. The use of this model allowed for obtaining key qualitative correspondences between the simulation and the experiment: 1) the region of maximum resistance exhibits a deformation; 2) this region is shifted along the deformation axis; 3) oscillations are observed on different sides of the axis depending on the polarity. These correspondences support the validity of the proposed explanation.

The repeatability of the polarity-dependent splitting of the oscillations, as shown in Fig. 2, was confirmed across several dozen MTJs with a diameter of 1000 nm. Qualitative agreement was also observed for MTJs with diameters of 600 and 400 nm, as shown in *Section S.5*.

Previously, the effect of vortex auto-oscillations under the influence of a spin current with vortex polarization was observed experimentally in spin valves [43] [44] [45] [46]. Although the influence of vortex chiralities and polarities was investigated, neither of the vortices was fixed. As a result, both vortices could shift under the influence of both the external field and the current passed through. This mobility of the structure prevented systematic studies of the effect. In previous theoretical works, an analytical description of the auto-oscillation process was provided through the energy description and the Tillier equation [40] [41] [42]. In this work, a phenomenological description is proposed, qualitatively explaining the reasons for the observed effect.

## 5. Conclusion

The paper experimentally demonstrates a method for the reversible magnetic reconfiguration of the synthetic antiferromagnet structure in MTJ using a voltage pulse. This mechanism allows for an analogous change in the residual resistance of the MTJ, which is non-volatile. Fixing the vortex state in the reference layer leads to the injection of a spin current with a vortex polarization distribution into the free layer. In this case, stable gyrotropic oscillations were experimentally observed in the absence of external fields for MTJ with diameter ranging from 400 nm to 1000 nm. The dynamics of the system were shown to exhibit a clear vortex core polarity symmetry breaking along an axis which was related to the history of the antiferromagnet reannealing procedure. Micromagnetic simulations well reproduced this effect by assuming a resultant deformation in the pinning antiferromagnet and pinned and reference layers. The ability to reconfigure individual MTJs both in terms of the static and dynamic properties after integration is significant interest to a range of potential applications, most specifically radio frequency spintronic neuromorphic computing.

## 6. Materials and Methods

The structures under study were obtained based on a film of the following composition from bottom to top: [Ta(5)/CuN(25)]x6/Ta(5)/Ru(5)/IrMn(6)/CoFe30(2)/Ru(0.7)/CoFe40B20(2.6)/MgO(5Ohm.um 2)/CoFe40B20(2)/Ta(0.21)/NiFe(7)/Ta(10)/Ru(7)/TiWN(15)/AlSiCu(200)/TiWN(15), thicknesses in nm. The film was deposited on the surface of the 200 mm thermally oxidized wafer Si/SiO2(200 nm) using magnetron sputtering Singulus TIMARIS Multi-Target-Module. The layers below the first IrMn layer are further used to form the bottom electrical contact and are optimized to minimize roughness. The layers above the NiFe layer are needed to protect the magnetic stack from numerous further etching procedures. Then, nano-pillars with a diameter from 100 to 1000 nm were formed using electron beam lithography and ion beam milling. The nano-pillar was insulated with SiO2, after which the bottom and top electrical contacts were formed, as shown in Fig.1a. The thickness of the MgO layer was selected to obtain RxA=5 Ωμm2. In this case, the tunnel magnetoresistance is 150% at a reading current of 0.1 mA. When the reading current increases to 10 mA, the value decreases to 80%. After fabrication, the structures were annealed at 330 C for 2 hours in the presence of $B_x$ = 1 T, to fix the magnetization orientation in the reference layer due to exchange bias effect at the interface with IrMn.

The magnetic properties of the stack were studied using a vibrating sample magnetometer (VSM) equipped with a thermal setup on a separate sample with an area of 6×10 $mm^2$. The magnetotransport properties were investigated using a probe station equipped with an electromagnet generating an in-plane magnetic field $B_y$ with an amplitude of up to 0.1 T and a two-point high-frequency probe. To generate a magnetic field along the x-axis, a field line fabricated over the MTJ column was used (Fig. 1a), enabling the creation of magnetic fields up to 40 mT with an efficiency of 0.2 mT/mA. The hysteresis loop was recorded using a Keysight B2901B source measurement unit (SMU), which measured the resistance of the structure during magnetic field sweeps along the x- or y-axis. Auto-oscillations were detected using an Agilent E446A spectrum analyzer and an Agilent DSO-X 92004A oscilloscope, both connected through a bias-tee (Fig. 1a). The magnetic reconfiguration procedure was performed by applying pulses from the same SMU in the presence of a constant magnetic field generated by the field line and/or electromagnet.

## 7. Acknowledgments

This work has received funding from the European Union's Horizon 2020 research and innovation programme under grant agreement No 101017098 (project RadioSpin), No 899559 (project SpinAge) and No 101070287 (project Swan-on-chip), Horizon Europe under grant agreement n°101070290 (NIMFEIA).

## Supplementary Information

### S.1. Magnetic Properties Of The Film

The film with the composition IrMn(6)/CoFe30(2)/Ru(0.7)/CoFe40B20(2.6)/MgO/CoFe40B20(2)/Ta(0.21)/NiFe(7) was annealed for two hours at a temperature of 330 °C in an in-plane magnetic field of 1 T to fix the single-domain state in a specific direction in both the reference and pinning layers (Fig. S1a). Figure S1b shows a representative hysteresis loop for the film, measured using a vibrating sample magnetometer (VSM) at room temperature. Key stages of magnetization reversal are marked on the loop. In state (1), the magnetization in all layers is aligned with the external negative magnetic field. During the transition from state (1) to state (2), the magnetization in the pinned layer rotates due to the combined effects of exchange bias at the interface with the antiferromagnetic (AFM) layer and indirect antiferromagnetic exchange interaction with the reference layer, which possesses a higher magnetic moment. An increase in the applied field in the positive direction causes switching of the free layer (state 3). Its hysteresis loop is shifted due to weak interactions with the reference layer, including indirect exchange or Néel coupling (Fig. S1c). A further increase in the field leads to reversal of the magnetization in the reference layer (state 4), although this reversal is hindered by the indirect exchange interaction with the pinned layer.

Based on the hysteresis loop, the saturation magnetization (Ms) and indirect exchange energy ($J_{ex}$) for each layer were calculated [53], as shown in Fig. S1b. The saturation magnetization is 0.66 MA/m for the free layer, 1.02 MA/m for the reference layer, and 0.76 MA/m for the pinning layer. The indirect exchange energy between the pinning and reference layers is –0.3 mJ, and –6 μJ between the reference and free layers.

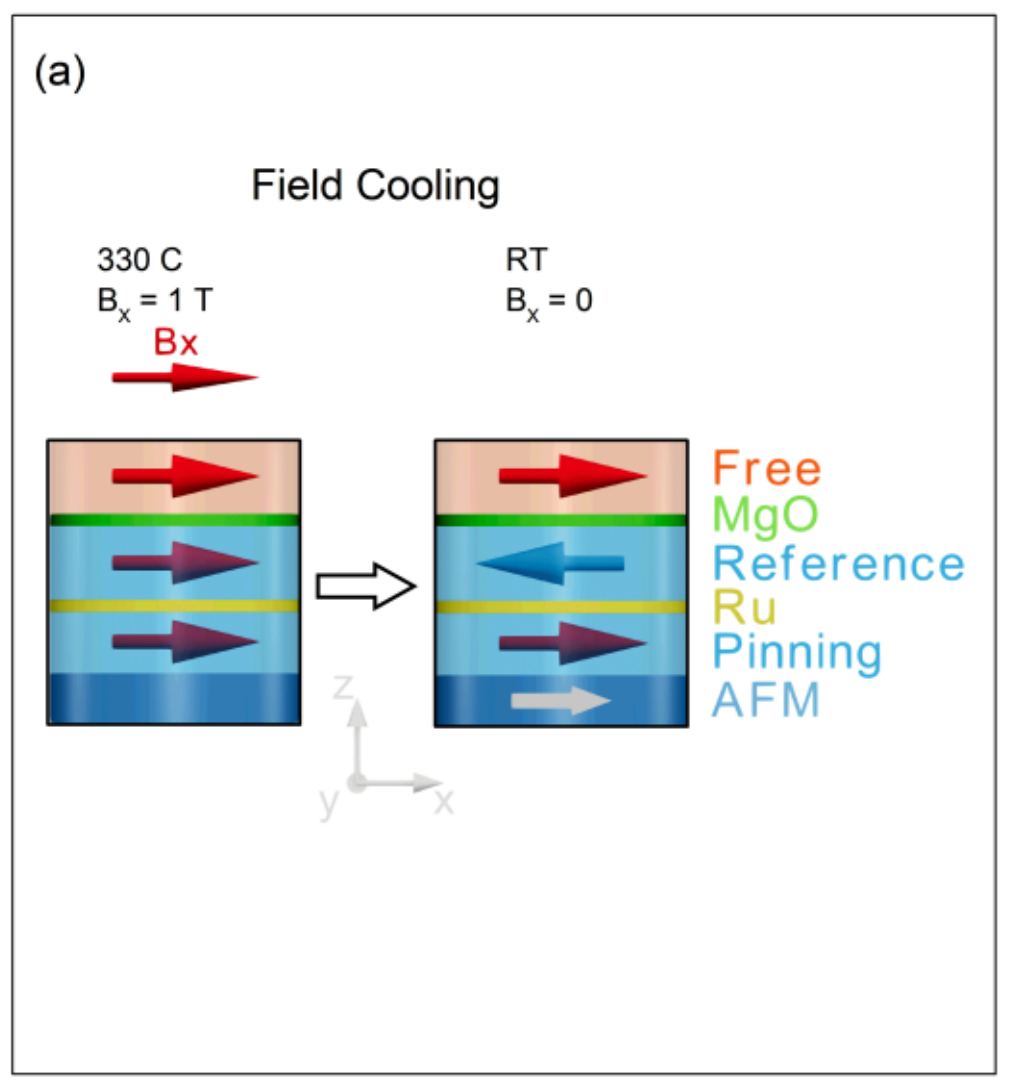

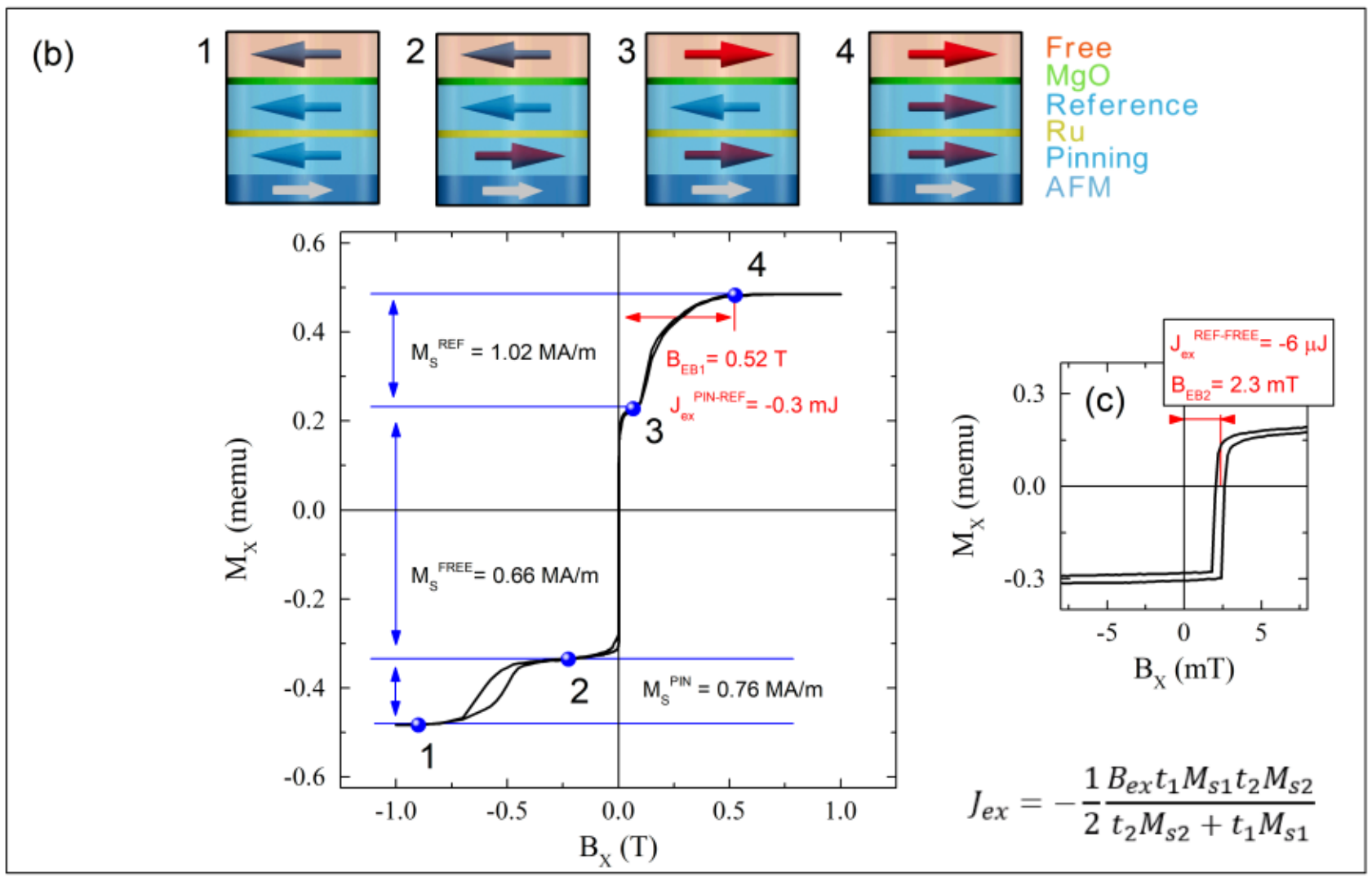

*Fig.S1. (a) Field cooling procedure. The sample was annealed at 330°C for three hours in a 1 T field along the x-axis and then cooled in the same field. (b) The magnetic hysteresis loop was obtained from a continuous film 6 × 8 mm using a vibrating sample magnetometer. The inset shows a magnified view of the low-field region. The numbers indicating the corresponding stages of layer switching.*

## S.2. Neel Temperature Estimation

The reconfiguration of the reference layer presented in this work involves current-assisted heating of the AFM layer above its Néel temperature. To experimentally verify the values of this temperature, the following experiment was conducted. The film was examined using a vibrating sample magnetometer (VSM) equipped with a thermal stage. The sample was first heated to a specific temperature and then cooled in the presence of a +50 mT magnetic field. A hysteresis loop was then recorded at room temperature. The procedure was repeated, but with the sample cooled in the opposite field direction (–50 mT) after heating to the same temperature. Another hysteresis loop was then recorded at room temperature.

Figure S2(a–d) presents pairs of hysteresis loops obtained after heating the sample to temperatures ranging from 160 °C to 220 °C. It can be observed that field cooling at 200 °C allows the direction of the exchange bias to be switched between opposite states, as schematically illustrated in Fig. S2d. The temperature dependence of the field shift associated with the pinning layer switching is shown in Fig. S2c. A complete reversal of the magnetization orientation in the reference layer is observed at a temperature of 220 °C, which can be considered equivalent to the Néel temperature.

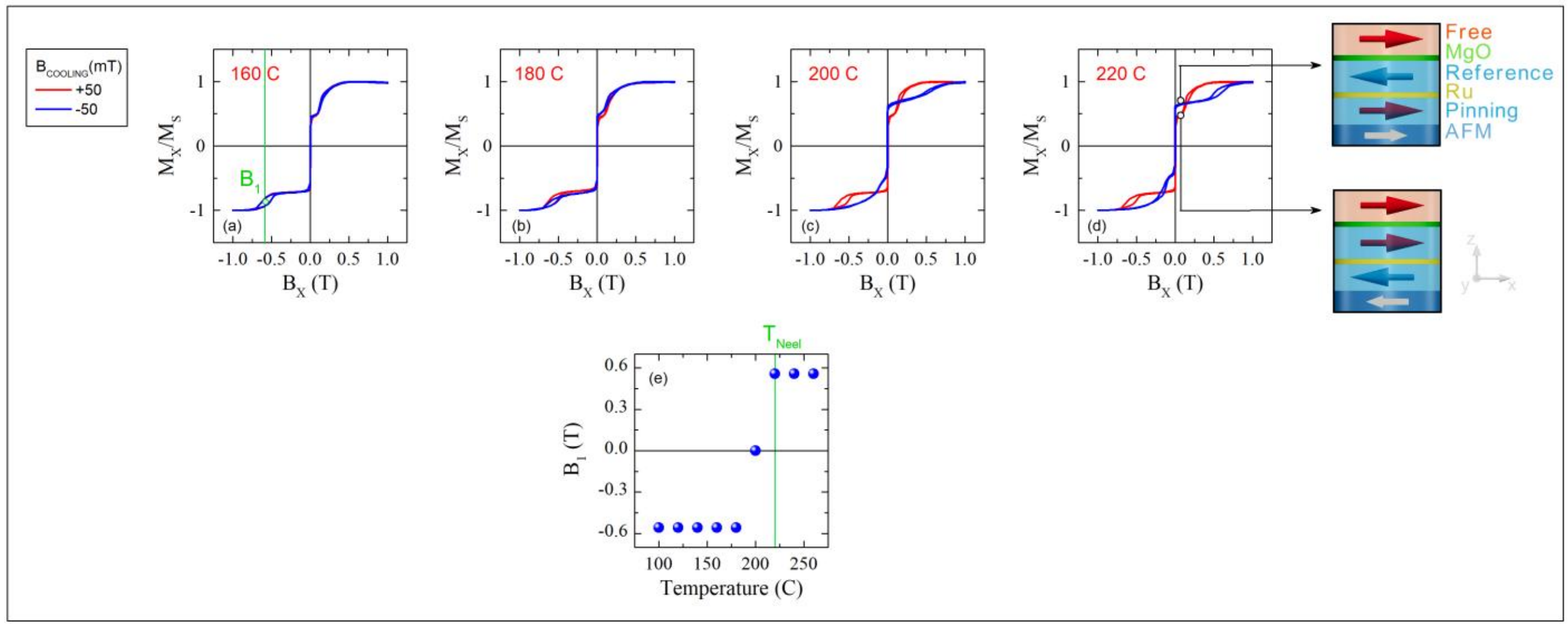

*Fig.S2. The field cooling procedure was performed in the thermal attachment of the vibrating sample magnetometer sequentially in the presence of +50 mT and -50 mT fields. The corresponding pair of hysteresis loops is shown in graphs (a-d) for different heating temperatures. Schematic illustrations of the structure depict the switching of the fixed magnetization orientation in the pinning and reference layers. Plot (e) shows the temperature dependence of field B1, which is indicated in plot (a).*

## S.3. Oersted Field Estimation

The propagation of a current pulse promotes the formation of a magnetic vortex structure in the reference layer due to the generation of an Oersted field. For a quantitative estimation of the relationship between the current and the resulting magnetic field, an analytical model was used. Within the conductor, the field distribution follows a circular symmetry, as described by relation $B = \frac{\mu_0 r}{2\pi R^2}$, which is obtained from the Biot–Savart law. The amplitude of the magnetic field remains constant at a fixed radius, and its orientation at each point is tangential to the corresponding circle (Fig. S3a).

The Oersted field is zero at the center of the MTJ column and increases linearly toward the periphery (Fig. S3b). For a column with a diameter of 1000 nm, the maximum field reaches approximately 4 mT for a 10 mA current, which is used to excite oscillations, and 24 mT for a 60 mA current, which is used for reconfiguration (Fig. S3b). Approximate values of the Oersted field at the periphery of columns with different diameters are presented in Fig. S3c, based on experimental data.

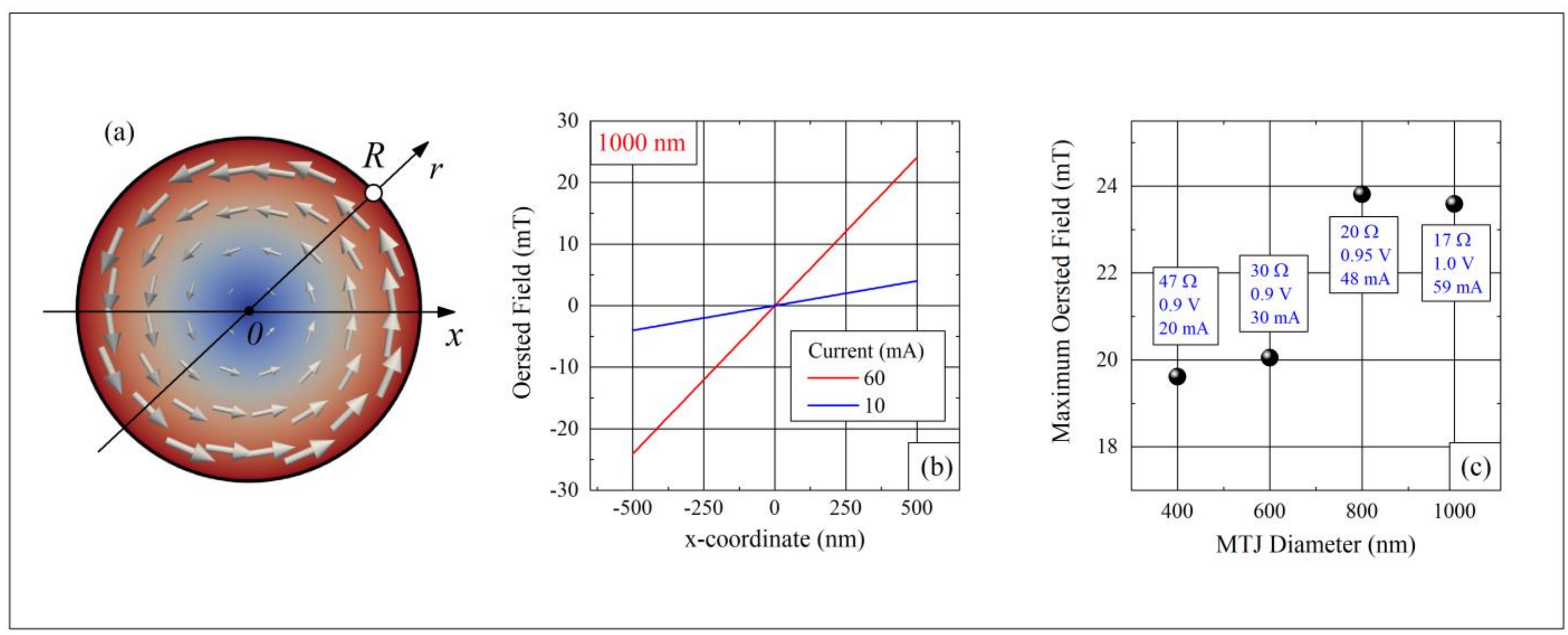

*Fig.S3. (a) Schematic diagram of the Oersted field distribution in the disk when a current is passed. (b) Calculated dependence of the Oersted field in a disk with a diameter of 1000 nm*

*on the x-coordinate for two different current values. (c) Calculated maximum Oersted field achieved during reconfiguration at the periphery of the disk for different diameters, taking into account the experimentally obtained parameters of the structures.*

## S.4. Oscillations for Different SAF States

The main text of the article demonstrates the relationship between the position of the fixed vortex in the reference layer and the auto-oscillation region using two-dimensional field scans (Fig. 2). In this example, the vortex was fixed at three different positions along the Y-axis. To achieve this, a magnetic field was applied along the X-axis during the reconfiguration process. We also experimentally demonstrated that by simultaneously applying a field along the X and Y-axis, it is possible to fix the vortex core in a controlled position, shifted along both the X- and Y-axes (Fig. S4). In this case, the oscillation regions shift accordingly in the two-dimensional scans.

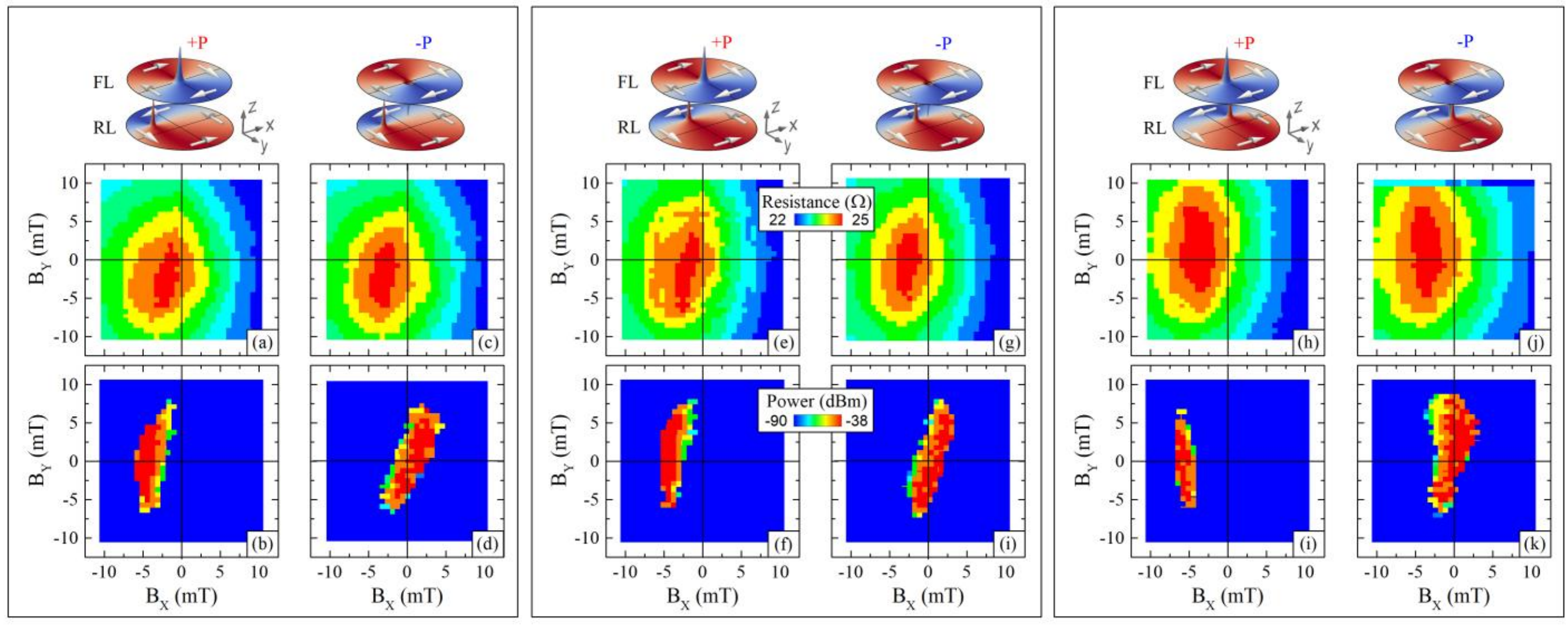

Fig.S4. *Distribution of MTJ resistance and oscillation power (diameter: 1000 nm) as a function of the $B_x$ and $B_y$ fields, obtained for different polarities of the vortex core in the free layer (i.e., +P, -P). In all cases, a constant current of +8 mA was applied. Diagrams (a-d) were obtained as a result of reconfiguration in the presence of $+B_x$ and $+B_y$. Diagrams (e-i) were obtained from reconfiguration in the presence of $+B_x$ and $B_y = 0$. Diagrams (h-k) were obtained from reconfiguration in the presence of $+B_x$ and $-B_y$.*

## S.5. Oscillations for Different MTJ diameters

The main text of the article is dedicated to describing the auto-oscillation and reconfiguration processes observed in a 1000 nm diameter MTJ. Nevertheless, all conclusions remain valid for devices with smaller diameters, as confirmed experimentally. The reconfiguration procedure enables the creation of a vortex in the reference layer in devices with diameters as small as 400 nm. For devices with diameters of 400, 600, and 800 nm, the process allows for the selection of appropriate field configurations along the X and Y axes, enabling auto-oscillations in the absence of an external field (Fig. S5).

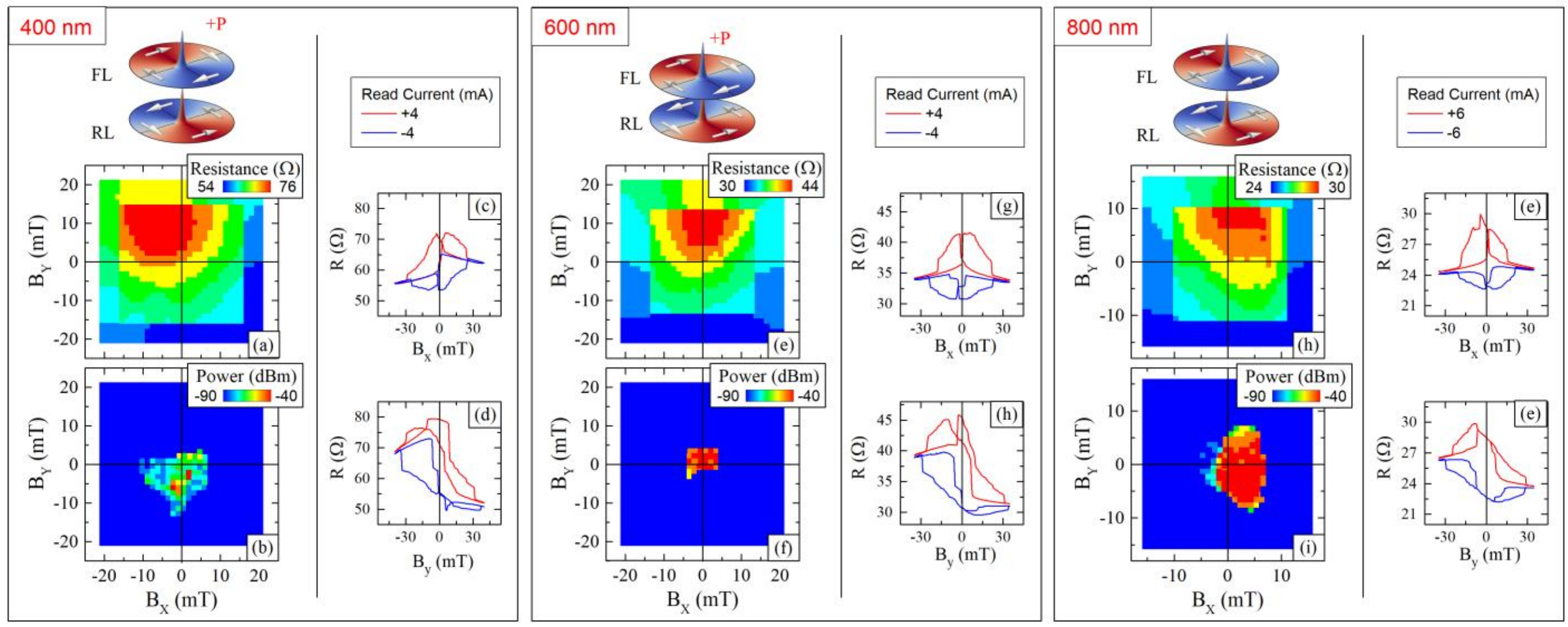

*Fig.S5.1. Distribution of MTJ resistance and oscillation power as a function of the $B_x$ and $B_y$ fields, obtained for different polarities of the vortex core in the free layer (i.e., +P, -P) for different diameter of MTJ. In all cases, zero-field self-oscillations are obtained.*

Fig. S5.2 summarizes the current dependence for disks of various diameters in a state that supports auto-oscillations in the absence of an external field. As the current increases, the MTJ resistance decreases, partly due to vortex oscillations in the free layer. In all cases, the oscillation frequency increases with current, while the oscillation power rapidly reaches saturation.

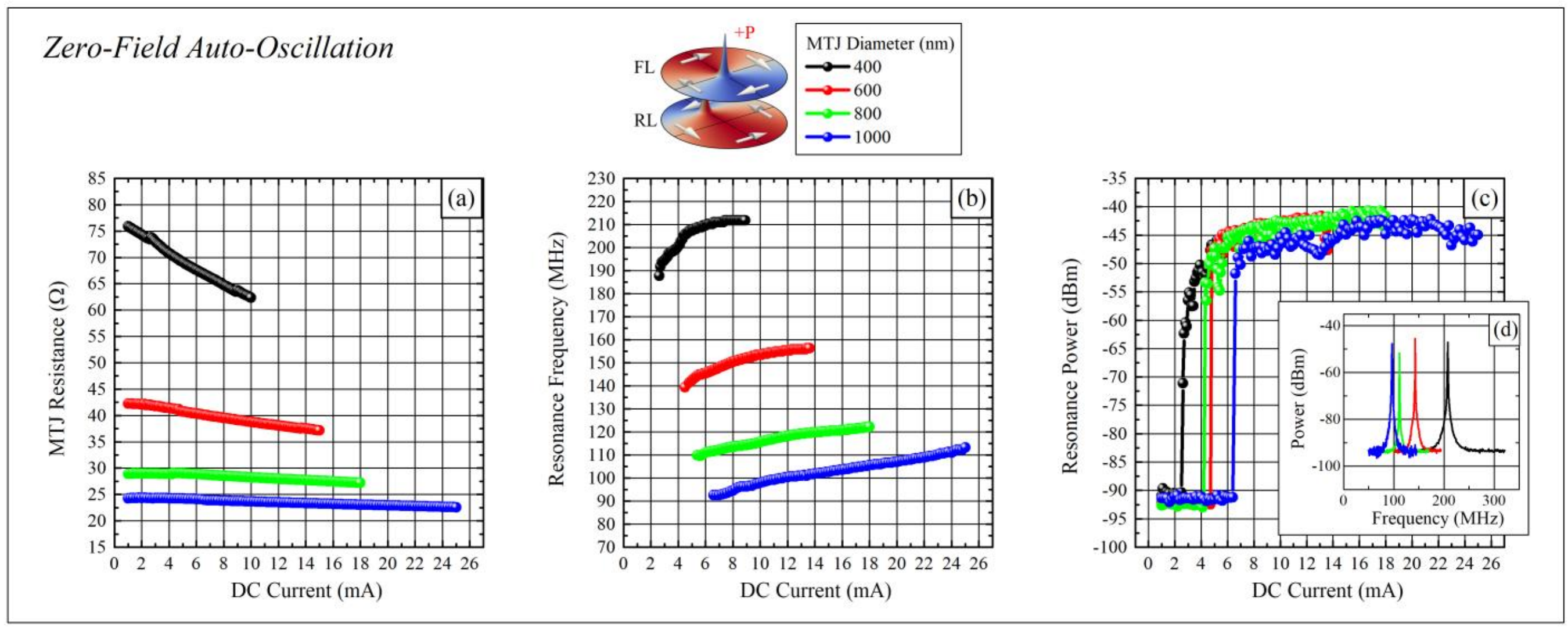

*Fig.S5.2. Results of experimental study of auto-oscillations in zero field for structures of diameter 400, 600, 800 and 1000 nm. Dependence of structure resistance (a) frequency (b) and reflected power (c) of oscillations on the intensity of the passed current. (d) Examples of frequency spectra for a current of 7 mA.*

## S.6. The SAF Energy Dependence On Magnetic States

The formation of magnetic vortices in antiferromagnetically coupled reference and pinning layers is a complex process, even when the exchange bias is switched off. Therefore, in this work, the possibility of vortex formation is considered from the perspective of system energy minimization. Using MuMax3, the energy density of a system consisting of two disks was estimated for configurations with either fixed magnetic vortices or single-domain states with antiferromagnetic alignment. Experimentally obtained values of the saturation magnetization and exchange interaction energy were used (see Section S.1). In the absence of an external magnetic field, the minimum energy configuration corresponds to single-domain states across all diameters. However, the presence of the Oersted field (Fig. S3a) makes vortex formation energetically favourable (Fig. S6). Since the saturation magnetization of the reference layer is higher than that of the pinning layer, the vortex in the reference layer repeats the chirality of the Oersted field. The magnitude of the Oersted field is directly proportional to the applied current, the dependence is plotted in milliamperes (mA) for ease of comparison with experimental data.

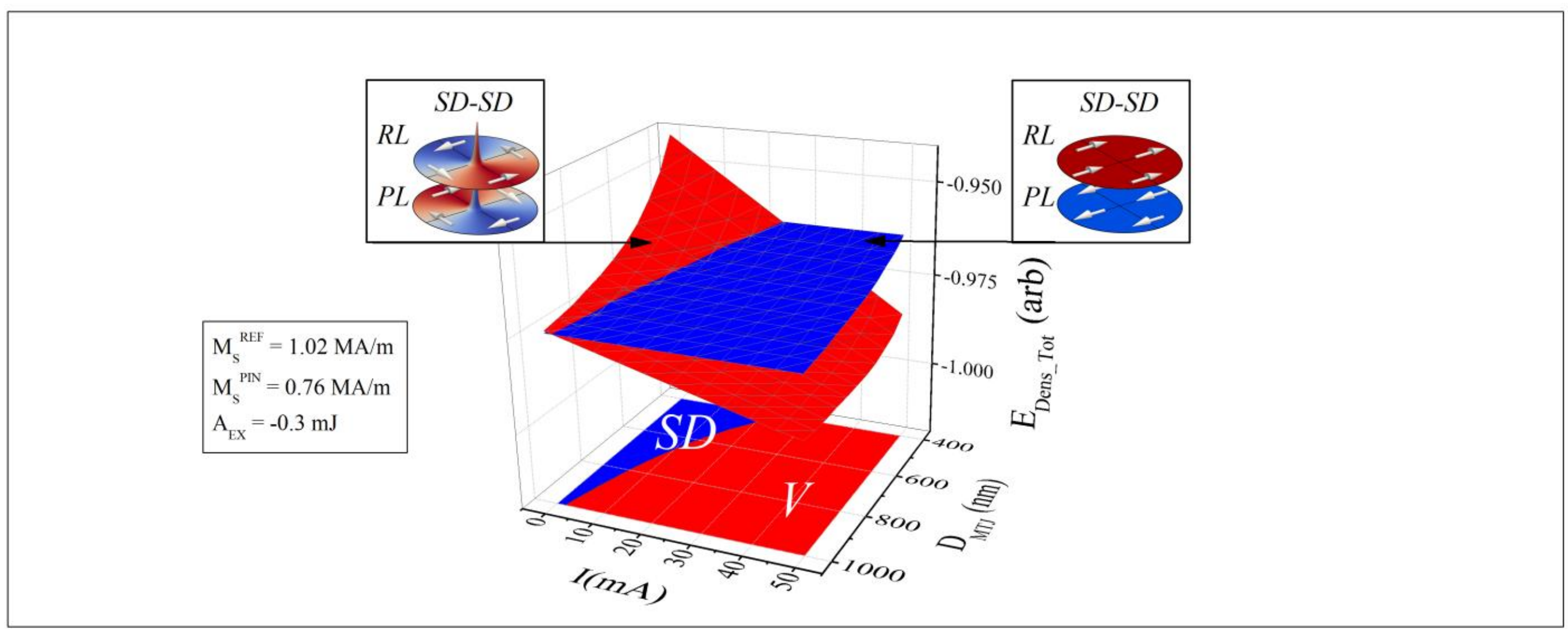

*Fig.S6. The dependences of the total energy of antiferromagnetically exchange-coupled layers on the diameter of the disks and the current passed through them, which generates the Oersted field. The dependences are obtained for the case of a single-domain statutes (SD-SD) in layers and a vortex structure (V-V).*

## S.7. Impact of RL Magnetic State on the Resistance Distribution

During the two-dimensional scanning of the external field along the x- and y-axes, the vortex in the free layer shifts relative to the fixed magnetic structure of the reference layer. The resulting resistance distribution allows us to make indirect assumptions about the fixed magnetic structure. To confirm the assumptions made, experimental results are compared with the simulation results.

Tunnel magnetoresistance is directly proportional to the cosine [54] of the angle between the magnetic moments in the free and reference layers, Fig. 1. Using the results of micromagnetic modeling, the resistance was calculated for each pair of cells in the free and reference layers. Then these values were summed up and divided by the number of cells. As a result, the resistance value was obtained for a specific pair of magnetic states in the free and reference layers.

To construct the resistance distribution as a function of external fields, magnetic states of the vortex in the free layer were simulated at field values Bx and By ranging from ±5 mT, resulting in 400 distinct states. For each state, the resistance was calculated based on the magnetic configuration fixed in the reference layer. This produced resistance distributions corresponding to various recorded configurations in the reference layer (Fig. 1).

When a symmetric vortex is fixed, the resistance distribution is also strictly symmetrical (Case 1). Deformation of the vortex along a single axis results in an elliptical resistance distribution (Case 2). Asymmetric deformation of the vortex in the reference layer produces a resistance distribution that qualitatively matches those observed experimentally (Case 3). If a discontinuity line is formed, the distribution no longer exhibits a localized resistance maximum (Cases 4 and 5). It is presumed that such cases may occur in small-diameter disks.

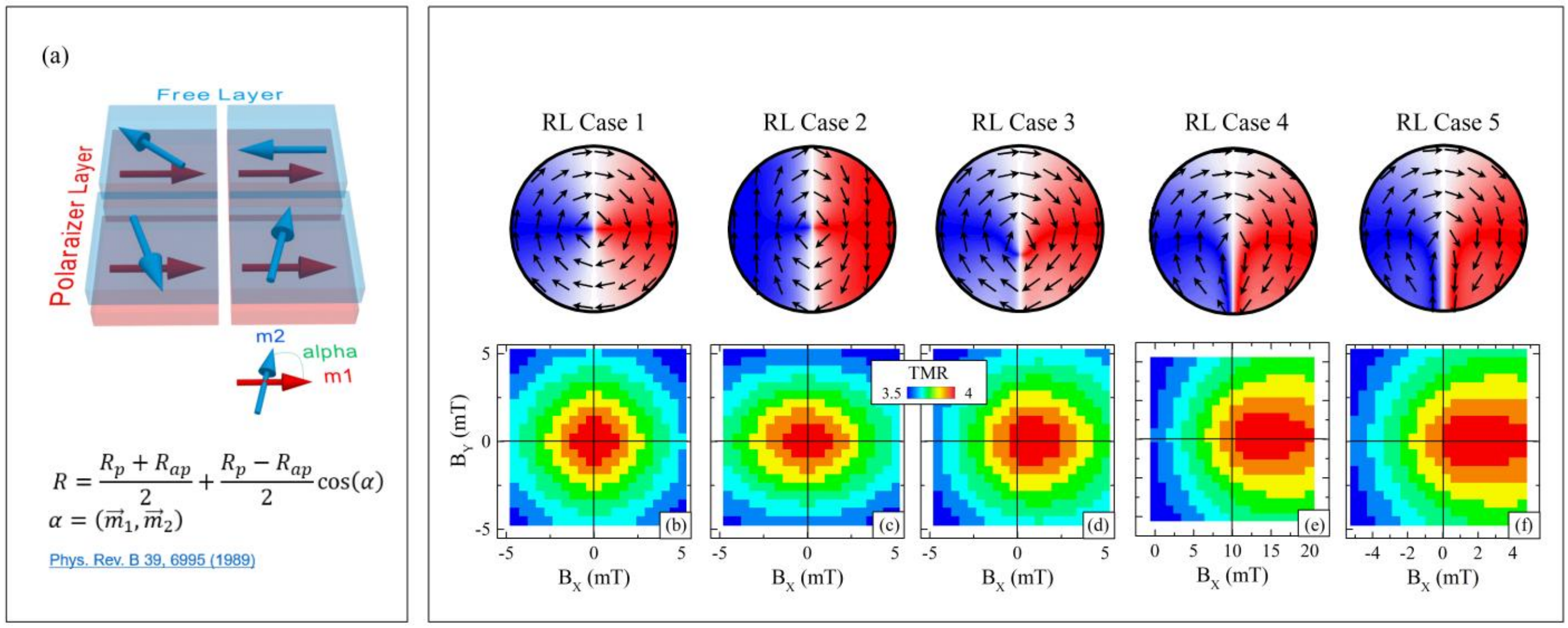

*Fig.S7. (a) Schematic description of the method used to calculate the magnitude of resistance change due to the tunnel magnetoresistance effect in a structure divided into cells. (b) Examples of resistance distribution in a structure where the reference layer has a fixed magnetic vortex with different deformation (b-f), and the symmetrical magnetic vortex with opposite chirality in the free layer shifts depending on the applied Bx and By fields.*

## Bibliography

[1] Hirohataa, A., Yamadab, K., Hillebrandse, B., "Review on spintronics: Principles and device applications," *Journal of Magnetism and Magnetic Materials,* vol. 509 , p. 166711, 2020.

[2] "Spintronics for ultra-low-power circuits," *Nat Rev Electr Eng,* vol. 1, p. 691, 2024.

[3] Marrows, Ch., Joseph Barker, J., Moorsom, T., "Neuromorphic computing with spintronics," *npj Spintronics,* p. 12, 2024.

[4] Roy, K., Cheng Wang, Ch., Sengupta, A., "Spintronic neural systems," *Nature Reviews Electrical Engineering,* vol. 1, p. 714, 2024 .

[5] Böhnert, T., Rezaeiyan, Y., Ferreira, R., "Weighted spin torque nano-oscillator system for neuromorphic computing," *Communications Engineering* , vol. 2, p. 65, 2023.

[6] Louis, S., Tyberkevych, V., Slavin, A., "Low Power Microwave Signal Detection With a Spin-Torque Nano-Oscillator in the Active Self-Oscillating Regime," *IEEE Transactions on Magnetics,* vol. 53, p. 1400804, 2017.

[7] Msiska, R., Love, J., Everschor-Sitte, K., "Audio Classification with Skyrmion Reservoirs," *Adv. Intell. Syst.,* vol. 5, no. 2, p. 2200388, 2023.

[8] Ross, A., Leroux, N., Grollier, J., "Multilayer spintronic neural networks with radiofrequency connections," *Nature Nanotechnology,* vol. 18, p. 1273, 2023 .

[9] "A crossbar array of magnetoresistive memory devices for in-memory computing," *Nature,* vol. 601, p. 211, 2022.

[10] "Microwave artificial neurons based on magnetic tunnel junction nano-oscillators for image recognition and denoising," *Japanese Journal of Applied Physics,* vol. 63, p. 100904 , 2024.

[11] Romera, M., Talatchian, Ph., Grollier, J., "Vowel recognition with four coupled spin-torque nano-oscillators," *Nature,* vol. 563 , p. 230, 2018.

[12] Zhu, J.-G., Park, Ch., "Magnetic tunnel junctions," *Materials Today,* vol. 9, p. 36, 2006.

[13] Bhatti, S., Rachid Sbiaa, R., Piramanayagam, S., "Spintronics based random access memory: a review," *Materials Today,* vol. 20, p. 530, 2017.

[14] Maciel, N., Marques, E., Cai, H., "Magnetic Tunnel Junction Applications," *Sensors,* vol. 20, p. 121, 2020.

[15] Skirdkov, P., Zvezdin, K., "Spin-Torque Diodes: From Fundamental Researchto Applications," *Ann. Phys. (Berlin),* vol. 532, p. 1900460, 2020.

[16] Fang, B., Carpentieri, M., Zeng, Z., "Giant spin-torque diode sensitivity in the absence of bias magnetic field," *Nature Communications,* vol. 7, p. 11259, 2016.

[17] Jiang, Sh., Yao, L., Akerman, J., "Spin-torque nano-oscillators and their applications," *Appl. Phys. Rev.,* vol. 11, p. 041309 , 2024.

[18] Sharma, R., Mishra, R., Yang, H., "Electrically connected spin-torque oscillators array for 2.4 GHz WiFi band transmission and energy harvesting," *Nature Communications,* vol. 12, p. 2924, 2021.

[19] Fang, B., Carpentieri, M., Zeng, Z., "Experimental demonstration of spintronic broadband microwave detectors and their capability for powering nanodevices," *Phys. Rev. Appl.,* vol. 11, p. 014022 , 2019.

[20] Sharma, R., Sisodia, N., Muduli, P., "Enhanced Modulation Bandwidth of a Magnetic Tunnel Junction-Based Spin Torque Nano-Oscillator Under Strong Current Modulation," *IEEE Electron Device Letters,* vol. 42, p. 1886, 2021.

[21] Ruiz-Calaforra, A., Purbawati, A., Ebels, U., "Frequency shift keying by current modulation in a MTJ-based STNO with high data rate," *Appl. Phys. Lett.,* vol. 111, p. 082401 , 2017.

[22] Litvinenko, A., Iurchuk, V., Slavin, A., "Ultrafast sweep-tuned spectrum analyzer with temporal resolution based on a spin-torque nano-oscillator," *Nano Lett.,* vol. 20, p. 6104 , 2020.

[23] Fukushima, A., Yamamoto, T., Yuasa, Sh., "Recent progress in random number generator using voltage pulse-induced switching of nano-magnet: A perspective," *APL Mater.,* vol. 9, p. 030905 , 2021.

[24] Jenkins, A., Alvarez, L., Ferreira, R., "Nanoscale true random bit generator based on magnetic state transitions in magnetic tunnel junctions," *Scientific Reports,* vol. 9, p. 15661, 2019.

[25] Zeng, Z., Amiri, P., Jiang, H., "High-Power Coherent MicrowaveEmission from Magnetic TunnelJunction Nano-oscillators withPerpendicular Anisotropy," *ACS Nano,* vol. 6, p. 6115, 2012.

[26] Tamaru, S., Kubota, H., Fukushima, A., "Extremely Coherent Microwave Emission from Spin Torque Oscillator Stabilized by Phase Locked Loop," *Scientific Reports,* vol. 5, p. 18134, 2016.

[27] Zhou, S., Zheng, C., Liu, Y., "Skyrmion-based spin-torque nano-oscillator in synthetic antiferromagnetic nanodisks," *Journal of Applied Physics,* vol. 128, p. 033907, 2020.

[28] Zhou, Y., Iacocca, E., Åkerman, J., "Dynamically stabilized magnetic skyrmions," *Nature Communications,* vol. 6, p. 8193 , 2015.

[29] Lendınez, S., Statuto, N., Macia, F., "Observation of droplet soliton drift resonances in a spin-transfer-torque nanocontact to a ferromagnetic thin film," *Physical Review B,* vol. 92, p. 174426 , 2015.

[30] Dussaux, A., Georges, B., Fert, A., "Large microwave generation from current-driven magnetic vortex oscillators in magnetic tunnel junctions," *Nature Communications,* vol. 1, p. 8, 2010.

[31] Dussaux, A., Grimaldi, E., Fert, A., "Large amplitude spin torque vortex oscillations at zero external field using a perpendicular spin polarizer," *Appl. Phys. Lett.,* vol. 105, p. 022404 , 2014.

[32] Hu, H., Yu, G., Zhou, H., "Design of a Radial Vortex-Based Spin-Torque Nano-Oscillator in a Strain-Mediated Multiferroic Nanostructure for BFSK/BASK Applications," *Micromachines,* vol. 13, p. 1056, 2022.

[33] Zvezdina, K., Ekomasov, E., "Spin Currents and Nonlinear Dynamics of Vortex Spin Torque Nano-Oscillators," *Physics of Metals and Metallography,* vol. 123, p. 201, 2022.

[34] K. Y. Guslienko, "Magnetization reversal due to vortex nucleation, displacement, and annihilation in submicron ferromagnetic dot arrays," *PHYSICAL REVIEW B,* vol. 65, p. 024414, 2001.

[35] Gaididei, Y., Kravchuk, V., Sheka, D., "Magnetic Vortex Dynamics Induced by an Electrical Current," *International Journal of Quantum Chemistry,* vol. 110, p. 83, 2010.

[36] Torrejon, J., Mathieu Riou, M., Grollier J., "Neuromorphic computing with nanoscale spintronic oscillators," *Nature,* vol. 547, p. 428, 2017.

[37] Shreya, S., Jenkins, A., Farkhani, H., "Granular vortex spin-torque nano oscillator for reservoir computing," *Scientifc Reports,* vol. 13, p. 16722, 2023.

[38] Farcis, L., Teixeira, B., Buda-Prejbeanu, D., "Spiking Dynamics in Dual Free Layer Perpendicular Magnetic Tunnel Junctions," *Nano Letters,* vol. 23, p. 7869, 2023.

[39] Choi, Y.-S., Lee, K.-S., Kim, S.-K., "Quantitative understanding of magnetic vortex oscillations driven by spin-polarized out-of-plane dc current: Analytical and micromagnetic numerical study," *Physical Review B ,* vol. 79, p. 184424 , 2009.

[40] Khvalkovskiy, A., Grollier, J., Cros, V., "Nonuniformity of a planar polarizer for spin-transfer-induced vortex oscillations at zero field," *Applied Physics Letters,* vol. 96, p. 212507, 2010.

[41] Sluka, V., Kakay, A., Schneider, C., "Quenched Slonczewski windmill in spin-torque vortex oscillators," *Physical Review B,* vol. 86, p. 214422 , 2012.

[42] Carpentieri, M., Martinez, E., Finocchio, G., "High frequency spin-torque-oscillators with reduced perpendicular torque effect based on asymmetric vortex polarizer," *Journal of Applied Physics,* vol. 110, p. 093911 , 2011.

[43] Locatelli, N., Naletov, V., Fert, F., "Dynamics of two coupled vortices in a spin valve nanopillar excited by spin transfer torque," *Appl. Phys. Lett.,* vol. 98, p. 062501 , 2011.

[44] Sluka, V., Kakay, A., Herte, R., "Spin-torque-induced dynamics at fine-split frequencies in nano-oscillators with two stacked vortices," *Nature Communications,* vol. 6, p. 6409, 2015.

[45] Sluka, V., Deac, A., Schneider, C., "Spin-transfer torque induced vortex dynamics in Fe/Ag/Fe nanopillars," *Journal of Physics D: Applied Physics,* vol. 44 , p. 384002 , 2011.

[46] Locatelli, N., Lebrun, R., Cros, V., "Improved Spectral Stability in Spin-Transfer Nano-Oscillators: Single Vortex Versus Coupled Vortices Dynamics," *IEEE Transactions on Magnetics,* vol. 51, p. 4300206, 2015.

[47] Slonczewski, J., "Currents and torques in metallic magnetic multilayers," *Journal of Magnetism and Magnetic Materials,* vol. 247 , p. 324, 2002.

[48] Ralpha, D., Stilesb, M., "Spin transfer torques," *Journal of Magnetism and Magnetic Materials,* vol. 320 , p. 1190, 2008.

[49] Stebliy, M., Jenkins, A., Ferreira, R., "Non-Volatile Analog Control and Reconfiguration of a Vortex Nano-Oscillator Frequency," *Advanced Functional Materials,* p. 2405776, 2024.

[50] L. Lombard, "IrMn and FeMn blocking temperature dependence on heating pulse width," *J. Appl. Phys.,* vol. 107, p. 09D728 , 2010.

[51] Slonczewski, J., "Current-driven excitation of magnetic multilayers," *Journal of Magnetism and Magnetic Materials,* vol. 159, p. L1, 1996.

[52] Vansteenkiste, A., Leliaert, J., Waeyenberge, B., "The design and verification of MuMax3," *AIP Advances,* vol. 4, p. 107133 , 2014.